\documentclass[%
reprint,
superscriptaddress,
amsmath,amssymb,
pra,
]{revtex4-2}

\usepackage{graphicx}
\usepackage{dcolumn}
\usepackage{bm}
\usepackage[colorlinks,linkcolor=blue, urlcolor=blue, anchorcolor=blue, citecolor=blue]{hyperref}
\usepackage{subfigure}
\usepackage{physics}
\renewcommand{\tr}{\operatorname{Tr}}

\begin{document}
\title{Direct measurement of quantum Fisher information}
\author{Xingyu Zhang}
\affiliation{Zhejiang Province Key Laboratory of Quantum Technology and Device,
Department of Physics, Zhejiang University, Hangzhou, Zhejiang 310027,
China}
\author{Xiao-Ming Lu}
\affiliation{School of Sciences, Hangzhou Dianzi University, Hangzhou 310018, China}
\author{Jing Liu}
\affiliation{MOE Key Laboratory of Fundamental Physical Quantities Measurement, 
	National Precise Gravity Measurement Facility, School of Physics, 
	Huazhong University of Science and Technology, Wuhan 430074, China}
\author{Wenkui Ding}
\affiliation{Beijing National Laboratory for Condensed Matter Physics,
	Institute of Physics, Chinese Academy of Sciences, Beijing 100190, China}
\author{Xiaoguang Wang}
\email{xgwang1208@zju.edu.cn}

\affiliation{Zhejiang Province Key Laboratory of Quantum Technology and Device,
Department of Physics, Zhejiang University, Hangzhou, Zhejiang 310027,
China}
\begin{abstract}
In the adiabatic perturbation theory, Berry curvature is related to the generalized force, and the quantum metric tensor is linked with energy fluctuation. While the former is tested with numerous numerical results and experimental realizations, the latter is less considered. Quantum Fisher information, key to quantum precision measurement, is four times quantum metric tensor. It is difficult to relate the quantum Fisher information with some physical observable. One interesting candidate is square of the symmetric logarithmic derivative, which is usually tough to obtain both theoretically and experimentally. The adiabatic perturbation theory enlightens us to measure the energy fluctuation to directly extract the quantum Fisher information. In this article, we first adopt an alternative way to derive the link of energy fluctuation to the quantum Fisher information. Then we numerically testify the direct extraction of the quantum Fisher information based on adiabatic perturbation in two-level systems and simulate the experimental realization in nitrogen-vacancy center with experimentally practical parameters. Statistical models such as transverse field Ising model and Heisenberg spin chains are also discussed to compare with the analytical result and show the level crossing respectively. Our discussion will provide a new practical scheme to measure the quantum Fisher information, and will also benefit the quantum precision measurement and the understand of the quantum Fisher information.
\end{abstract}
\maketitle

\section{Introduction}
Quantum Fisher information (QFI) and quantum Fisher information matrix (QFIM), are at the heart of quantum precision measurement theory \citep{RevModPhys.89.035002,Liu_2019,PhysRevA.82.042103,hauke2016measuring,holevo2011probabilistic,helstrom1969quantum,PhysRevA.82.042103}. The QFI is usually denoted by $F_\lambda$ and originated from the classical Fisher information, depicts how
much information a quantum state contains for the metrology of certain
parameter. The sensitivity of some parameter $\Delta^{2}\lambda$ is
lower bounded by the reciprocal of QFI, $1/F_\lambda$. As a result, larger
quantum Fisher information is required for better estimation of the
corresponding parameter. QFIM $F_{\mu\nu}$ is a generalization of QFI, and is crucial to the multi-parameter estimation. Although the QFI is crucial in the measurement theory, only recently a few schemes \citep{PhysRevB.97.201117,PhysRevLett.116.090801,10.1093/nsr/nwz193,PhysRevLett.122.210401,PhysRevLett.127.260501,PhysRevResearch.3.043122,PhysRevA.105.012210} are proposed for the direct extraction of the QFI.
Moreover, the calculation of the QFI is not that simple, and the expression
of the estimating state is even not known. This remains an interesting
problem in the quantum measurement. Inspired by the work \citep{gritsev2012dynamical}, we notice that the adiabatic perturbation may be a starting point for the measurement of QFI, and a physical quantity naturally emerges from that.

The quantum geometric tensor \citep{PhysRevB.81.245129,PhysRevLett.121.020401,provost1980riemannian,gianfrate2020measurement} consists of two important physical quantities, i.e., the Berry curvature corresponding to its imaginary part, and the quantum metric tensor corresponding to its real part. The former is an intrinsic topological quantity, and its integral gives the Chern number \citep{RevModPhys.82.1959,SecondChernnumber}. The latter is usually less considered in condensed matter theory, however, but it plays a better role in quantum precision measurement since it is exactly a quarter of QFIM. Since the Berry curvature is particular important in topological physics, it is useful to find a way to measure it directly \citep{PhysRevLett.117.015301,PhysRevB.89.045107,luu2018measurement,doi:10.1126/science.aad4568}. It has been realized that the Berry curvature emerges as the dynamical response in the nonadiabatic evolution.
The excellent tool is the adiabatic perturbation theory \citep{avron2012adiabatic,Avron_2011,DeGrandi2010,PhysRevA.78.052508,PhysRevA.90.022104,
	PhysRevLett.104.170406,KOLODRUBETZ20171,De_Grandi_2013},
which regards the quantum adiabatic approximation as the zeroth-order
case and depicts a perturbation extension in terms of the small parameter's
changing rate $v$ to correct the quantum adiabatic theorem. In the
work of Gritsev and Polkovnikov \citep{gritsev2012dynamical},
they used the result from adiabatic perturbation theory to calculate
the linear response of a slowly driven system and the Berry curvature
emerges due to the quench. This means
\begin{figure*}[ht]
	\centering
	\includegraphics[width=18cm,height=7cm]{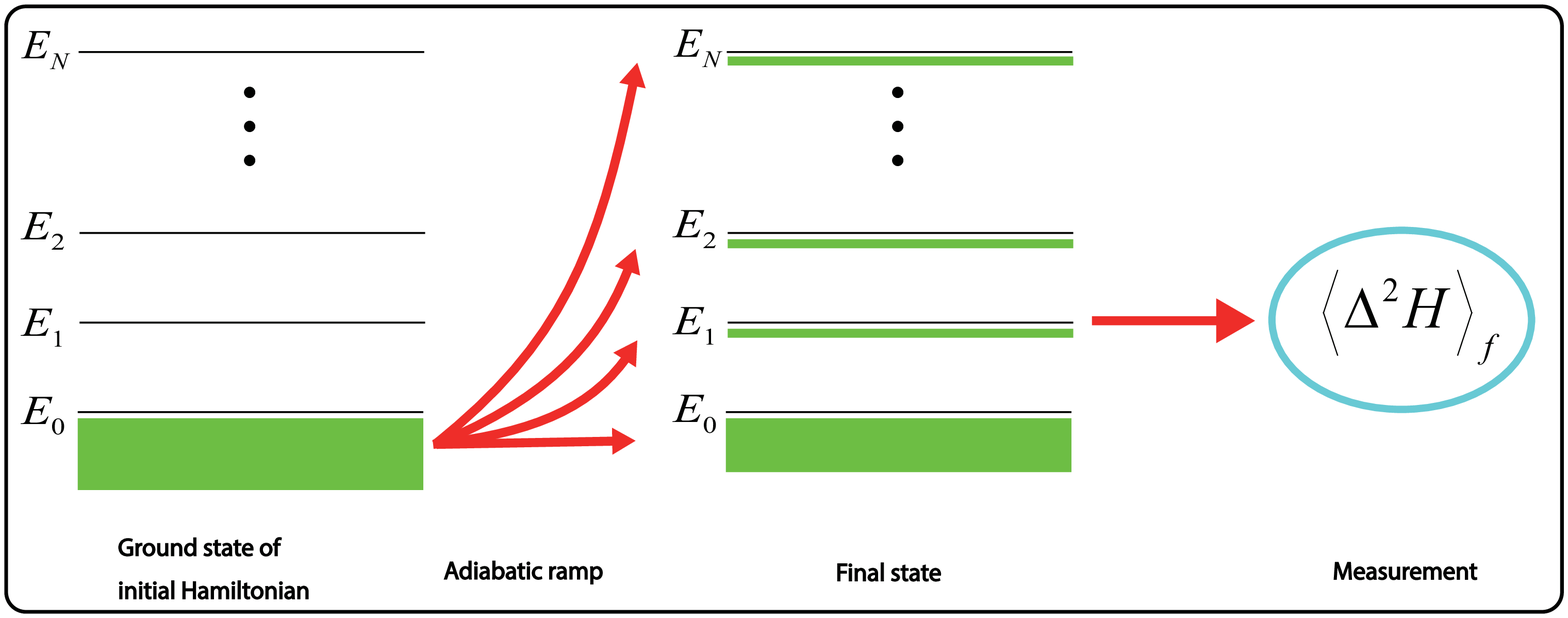}
	\caption{Illustration of the extraction of the QFI (matrix). The initial state is prepared to be the ground state of the initial Hamiltonian (The green block represents the population in each level and it is located in the ground state obviously). Then the parameter(s) of the Hamiltonian is (are) slowly changed with vanishing switching-on velocity. The populations in higher levels are still small but not zero due to quasi-adiabaticity. When the Hamiltonian is evolved into the interested one, the energy fluctuation is measured in the final state of the system. The ratio between the expectation to the square of the final velocity of parameter(s) gives the QFI (matrix).}
	\label{fig1}
\end{figure*}
\begin{equation}
	-\langle\psi(t_{f})|\partial_{\mu}H|\psi(t_{f})\rangle=\text{const}+\mathcal{B}_{\mu\lambda}v_{\lambda}+\mathcal{O}(v^{2}),\label{original}
\end{equation}
where the quench velocity $v_{\lambda}=\dot{{\lambda}}$ is the changing
rate of the parameter, and the above quantities are evaluated
in the final time $t_{f}$. The decorated $\mathcal{B}_{\mu\nu}$ is the Berry curvature of the ground state of the final Hamiltonian. The left-hand side of the equation is
called the generalized force. This remarkable finding can benefit
the direct measurement of the Berry curvature and even the topological
transition \citep{PhysRevLett.113.050402,PhysRevA.96.010101,PhysRevA.102.032613,
	PhysRevLett.127.136802}, regardless of the system's size and interaction strength.
On the other hand, in the work \citep{PhysRevB.88.064304,KOLODRUBETZ20171} they achieve a similar result concerning the quantum metric tensor, hence we find a way to directly measure the QFIM in the context of adiabatic perturbation. The results can be written as the following expressions with respect to QFIM:
\begin{eqnarray}
	\langle\psi(t_{f})|\Delta^2 H|\psi(t_{f})\rangle=\frac{1}{4}F_\lambda {v_\lambda}^2+\mathcal{O}(v^{3}),\label{second}\\
	\langle\psi(t_{f})|\Delta^2 H|\psi(t_{f})\rangle=\frac{1}{4}\sum_{\mu\nu}F_{\mu\nu} v^2+\mathcal{O}(v^{3}).\label{third}
\end{eqnarray}
$F_\lambda$ and $F_{\mu\nu}$ are the QFI and QFIM of the ground state of the final time Hamiltonian, respectively, because The ground state of a Hamiltonian is extremely important, since it exhibits the property of the system, e.g., quantum phase transition \citep{gu2010fidelity}. Only when two parameters are slowly driven with the same time-dependent part, the result gives the sum of QFIM entities (\ref{third}). These results show that the QFI of the ground state of the corresponding final Hamiltonian can be directly observed as the slope of the expectation value of energy fluctuation with respect to the square of the parameter's changing velocity. Therefore, we can measure the QFI of the ground state of nondegenerate Hamiltonians even if the ground state is not expressed explicitly. In contrast with Eq.~(\ref{original}), Eq.~(\ref{second}) and Eq.~(\ref{third}) are not considered much and not verified with numerical and experimental simulation. In this article, we depict how to use these results to extract the QFI and QFIM. Furthermore, experimental setup is considered with the nitrogen-vacancy center. 

This paper is divided into four parts. We first describe how to obtain Eq.~(\ref{second}) and Eq.~(\ref{third}) using a different way from that of \citep{PhysRevB.88.064304,KOLODRUBETZ20171}. Then we use a two-level system to demonstrate the validity of our method. Next, we briefly discuss the experimental protocols via the NV-center. Thirdly, we exhibit the applications of our result to the statistical models, i.e., one-dimensional transverse-field Ising model, to testify the validity and utility. Since the one-dimensional transverse-field Ising model can be solved analytically, so this provides a wonderful platform for us to compare our method with the analytic result. It is shown that our method matches the result of the standard procedure of diagonalization well. The last part concludes our work and give some discussions.

\section{Measuring the QFI of the ground state using adiabatic perturbation}
In this part, we will provide a method to measure the QFI and QFIM of certain
ground state based on the adiabatic perturbation. It is known that
for a Hamiltonian $H(\lambda(t))$, the spectrum can be expressed
as 
\begin{equation}
H(\lambda(t))|\phi_{n}(\lambda(t))\rangle=E_{n}(\lambda(t))|\phi_{n}(\lambda(t))\rangle,
\end{equation}
where $|\phi_{n}\rangle$ is the $n$th eigenstate of the Hamiltonian
$H$ with the eigenvalue $E_n$. We assume that the spectrum is finite and the Hamiltonian is nondegenerate.
The QFI of the ground state $|\phi_{0}\rangle$ of the Hamiltonian is 
\begin{align}
F_{\lambda} & =4(\langle\partial_{\lambda}\phi_{0}|\partial_{\lambda}\phi_{0}\rangle-\langle\partial_{\lambda}\phi_{0}|\phi_{0}\rangle\langle\phi_{0}|\partial_{\lambda}\phi_{0}\rangle)\nonumber \\
 & =4\sum_{n\ne0}\langle\partial_{\lambda}\phi_{0}|\phi_{n}\rangle\langle\phi_{n}|\partial_{\lambda}\phi_{0}\rangle\nonumber \\
 & =4\sum_{n\ne0}|\langle\partial_{\lambda}\phi_{0}|\phi_{n}\rangle|^{2}.\label{eq:Fisher}
\end{align}
To directly measure the QFI, we need to connect it with some
observables. 

In contrast with the method of \citep{PhysRevB.88.064304,KOLODRUBETZ20171}, we give alternative approach to prove Eq.~(\ref{second}) and (\ref{third}). First, we follow the work of Berry and Robbins~\citep{berry1993chaotic} to do the adiabatic expansion.
By using a small constant \(\epsilon\), called the adiabatic parameter, to change the time scale as \(t\to t /\epsilon\), the Schr\"odinger equation becomes
\begin{equation}
	i \epsilon \dv{\rho(t)}{t} = [H(\lambda(t)),\rho(t)],
\end{equation}
where the time-dependent quantities are rewritten with the new time scale, i.e., \(\rho(t/\epsilon) \to \rho(t)\) and \(\lambda(t/\epsilon) \to \lambda(t)\).

Berry and Robbins~\citep{berry1993chaotic} expanded the density operator of the evolving state as
\begin{equation} \label{eq:expansion}
	\rho(t)=\sum_{r=0}^{\infty}\epsilon^{r}\rho_{r}(t),
\end{equation}
where the $r$th-order $\rho_{r}(t)$ is required to satisfy
\begin{align}
	[H,\rho_{0}] & =0, \label{eq:zero_order}\\
	[H,\rho_{r}] & =\mathrm{i}\dot{\rho}_{r-1} \quad \forall\, r>0. \label{eq:rth_order}
\end{align}
It is easy to verify from Eq.~(\ref{eq:zero_order}) and (\ref{eq:rth_order}) that \(\rho(t)\) given by Eq.~(\ref{eq:expansion}) satisfies the Schr\"odinger equation.
Let \(\ket{\phi_k}\) be the instantaneous eigenstate of \(H\) with the instantaneous eigenvalue \(E_k\).
It follows from Eq.~(\ref{eq:rth_order}) that the off-diagonal elements of \(\rho_r\) can be determined by the time derivative of \(\rho_{r-1}\) as~\cite{berry1993chaotic}
\begin{equation}\label{eq:off_diagonal}
	\langle\phi_{k}|\rho_{r}|\phi_{l}\rangle
	=\mathrm{i}\frac{\langle\phi_{k}|\dot{\rho}_{r-1}|\phi_{l}\rangle}{E_{k}-E_{l}}
	= \mathrm{i} \dot\lambda \frac{\langle\phi_{k}|\partial_\lambda{\rho}_{r-1}|\phi_{l}\rangle}{E_{k}-E_{l}}
\end{equation}
for all $k\ne l$. 
The diagonal elements of \(\rho_r\) should be determined by other conditions in addition to Eq.~(\ref{eq:zero_order}) and Eq.~(\ref{eq:rth_order}), as the sum of \(\rho_r\) and any other constant operator that is simultaneously diagonalizable with \(H\) still satisfies Eq.~(\ref{eq:rth_order}).

Assume that the system is initially in the ground state of Hamiltonian.
In such case, \(\rho_0\) is chosen as the adiabatic state \(\op{\phi_0}\), i.e., the instantaneous ground state at time \(t\), which obviously satisfies the condition given by Eq.~(\ref{eq:zero_order}). 
We shall investigate the adiabatic expansion of the expectation \(\tr[(H-E_{0})^{2}\rho] \):
\begin{align} \label{eq:expectation_expansion}
	\quad \tr[(H-E_{0})^{2}\rho] 
	&= \sum_{r=0}^\infty 
	\epsilon^r \tr[(H-E_{0})^{2}\rho_r].
\end{align}
We can use the basis constituted by the instantaneous eigenstates of \(H\) to represent the operators, that is 
\begin{align}\label{eq:H2}
	\quad \tr[(H-E_{0})^{2}\rho] 
	= \sum_{k>0} (E_k-E_{0})^{2} p_k
\end{align}
where \(p_k\) for \(k>0\) is the transition probability defined as
\begin{equation}
	p_k := \ev{\rho}{\phi_k}
	= \sum_{r=1}^\infty \epsilon^r \ev{\rho_r}{\phi_k}.
\end{equation}
Berry and Robbins~\cite{berry1993chaotic} used the pure state condition, \(\rho(t)=\rho(t)^2\), to determine the diagonal elements of \(\rho_r\).
Using the adiabatic expansion \(\rho = \sum_r \epsilon^r \rho_r\), the pure state condition can be written as 
\begin{equation}
	\sum_{r=0}^\infty \epsilon^r \rho_r = \sum_{r=0}^\infty \sum_{s=0}^\infty \epsilon^{r+s} \rho_r \rho_s.
\end{equation}
The zeroth approximation of the above equality is automatically holds as the adiabatic state \(\rho_0\) is a pure state.
It then follows from the first order approximation \(\rho_1 = \rho_0 \rho_1 + \rho_1\rho_0\) that 
\begin{equation}
	\ev{\rho_1}{\phi_k} = 2\delta_{0k} \ev{\rho_1}{\phi_k},
\end{equation}
meaning that \(\ev{\rho_1}{\phi_k}=0\).
To the second order approximation, the pure state condition implies that \(\rho_2 = \rho_0\rho_2 + \rho_1\rho_1+\rho_2\rho_0\).
For \(k\neq0\), it follows that
\begin{align} \label{eq:rho2}
	\ev{\rho_2}{\phi_k} &= \sum_l \mel{\phi_k}{\rho_1}{\phi_l} \mel{\phi_l}{\rho_1}{\phi_k}. 
\end{align}
Combining the condition Eq.~(\ref{eq:rth_order}) and the first order approximation of the pure state condition, it can be shown that
\begin{align} \label{eq:rho1}
	\mel{\phi_k}{\rho_1}{\phi_l} = 
	\begin{cases}
		\mathrm{i}\dot{\lambda}\frac{\langle\phi_{k}|\partial_{\lambda}\phi_{l}\rangle(\delta_{0l}-\delta_{0k})}{E_{k}-E_{l}}, &\text{ if } k\ne l,\\
		0, &\text{ if } k=l.
	\end{cases}
\end{align}
Substituting Eq.~(\ref{eq:rho2}) and Eq.~(\ref{eq:rho1}) into Eq.~(\ref{eq:H2}) and neglecting the higher order terms, we get
\begin{align}
	\tr[(H-E_{0})^{2}\rho] 
	& \approx \epsilon^2 \sum_{k\ne0}\langle\phi_{k}|\rho_{2}(t)|\phi_{k}\rangle(E_{k}-E_{0})^{2} \nonumber\\
	&= \epsilon^2 \dot\lambda^2 \sum_{k\ne0} |\ip{\phi_k}{\partial_\lambda\phi_0}|^2.\label{main}
\end{align}

Note that the adiabatic parameter \(\epsilon\) can be absorbed into the velocity of \(\lambda\) as \(\epsilon\dot\lambda \to \dot\lambda\) when we change to the original time scale.
We denote the quantity $(H-E_0)^2$ as square of the absolute Hamiltonian (SAH), since the eigenvalues of this quantity are independent of the choice of the zero of the energy. Since SAH is different from the energy fluctuation by high order negligible quantities, i.e.,
\begin{align}
	&\langle(H-E_0)^2\rangle-\langle(H-\langle H \rangle)^2\rangle\nonumber\\
	=&(\langle H\rangle-E_0)^2\nonumber\\
	=&\bigl(\sum_{n\neq0}|a_n(\lambda_f)|^2E_n\bigr)^2\sim\mathcal{O}(\dot{\lambda}^4),
\end{align}
and then we recover the result Eq.~(\ref{second}), as briefly discussed in \cite{PhysRevB.88.064304,KOLODRUBETZ20171}. From this expression, once we obtain the expectation of the $\Delta^{2}H$ or $(H-E_0)^2$
at the final time, the QFI of the ground state of the final Hamiltonian
$H$ can be seen from the proportional coefficient with respect to
the square of velocity $\dot{\lambda}_f^{2}$.

We can conclude the measurement procedure above as follows. We set
the estimating parameter of the Hamiltonian at arbitrary initial value
and the initial state at the ground state. Then we evolve the parameter
of the Hamiltonian to the required value, and the evolving velocity
needs to be very slow. At the final time, we measure the energy fluctuation or SAH in the instantaneous state. The ratio of expectation with respect to the square of final time parameter's changing velocity gives
a quarter of QFI of the ground state of the final time Hamiltonian.
We do not need to calculate the explicit expression of the final time
ground state.

We now move to the extraction of the multi-parameter QFIM. Inspired by the work of Ozawa \citep{PhysRevB.97.201117}, we use a two-parameter modulation to extract the multi-parameter QFIM. When the arbitrarily selected two parameters share the same time-dependent part, i.e., $\lambda_1=C+f(t)$ and $\lambda_2=D+f(t)$, Eq.~(\ref{eq:rho1}) becomes
\begin{equation}
	\langle\phi_{k}|\rho_{1}(t)|\phi_{l}\rangle =\mathrm{i}\dot{\lambda}\frac{\sum_\mu\langle\phi_{k}|\partial_{\mu}\phi_{l}\rangle(\delta_{0l}-\delta_{0k})}{E_{k}-E_{l}},\ (k\ne l)\label{rho1mul}
\end{equation}
and Eq.~(\ref{main}) is corrected as 
\begin{equation}
\mathrm{Tr}[(H-E_{0})^{2}\rho(t)]=\epsilon^{2}\dot{\lambda}^{2}\sum_{\mu\nu=1}^2\sum_{k\ne0}\langle\partial_{\mu}\phi_{0}|\phi_{k}\rangle\langle\phi_{k}|\partial_{\nu}\phi_{0}\rangle,\label{secondmain}
\end{equation}
where $\dot{\lambda}=\mathrm{d}f(t)/\mathrm{d}t$ is the velocity of both parameters. The ratio between the final expectation of the SAH and the parameters' velocity gives a quarter of the sum of QFIM elements, as in Eq.~(\ref{third}). Notably, $C$ and $D$ are time-independent constants, and their difference should be set to satisfy the difference between our anticipated final $\lambda_1,\lambda_2$. The time-dependent part $f(t)$ is not necessarily periodically small perturbation and constrained only by the adiabatic conditions. Since $F_{\lambda_1 \lambda_1}$ and $F_{\lambda_2 \lambda_2}$ can be extracted from Eq.~(\ref{main}), the off-diagonal multi-parameter QFI ${F_{\lambda_1 \lambda_2}}$ is easily obtained. Next, we will show the feasibility of the measurement protocol on QFI and QFIM in the context of adiabatic perturbation.

\begin{figure}[h]
\includegraphics[scale=0.5]{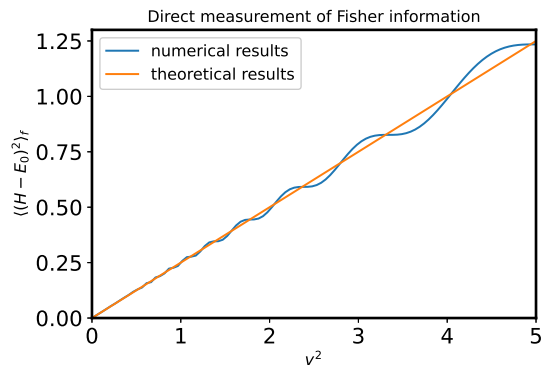}
\caption{The measurement of the QFI with respect to $\theta$ $F_\theta$ of the ground state of a two-level system.
	The numerical result is obtained by the dynamical simulation using the software packages developed in Refs.~\citep{JOHANSSON20121760,zhang2022quanestimation}.}
\label{QFI_twolevel}
\end{figure}

These results can be testified within a two-level system. We consider
the following typical Hamiltonian
\begin{equation}
H(t)=B\vec{n}(t)\cdot\vec{\sigma},\label{twoHam}
\end{equation}
where $\vec{n}(t)$ is a 3-component vectors defined as $\vec{n}(t)=(\sin\theta\cos\phi,\sin\theta\sin\phi,\cos\theta)$.
In this situation, we set the polar angle $\theta=\frac{v^{2}t^{2}}{2\pi}$
and the azimuthal angle $\phi=0$. The ground state of the initial
Hamiltonian is $(0,1)^{T}$, and it evolves with the time-dependent
Hamiltonian until the final time $t_f=\frac{\pi}{v}$. At the final
state, the observable $(H-E_{0})^{2}$ or $\Delta^2 H$ is measured. Since the parameter's
changing velocity at last is just $v$, the QFI of the ground state
of the final Hamiltonian $H$ is obtained. The theoretical
computation of the final QFI is unity and the above procedure gives
the same result from the Fig.~\ref{QFI_twolevel}.

\begin{figure}[h]
\centering
\subfigure[]{\includegraphics[scale=0.55]{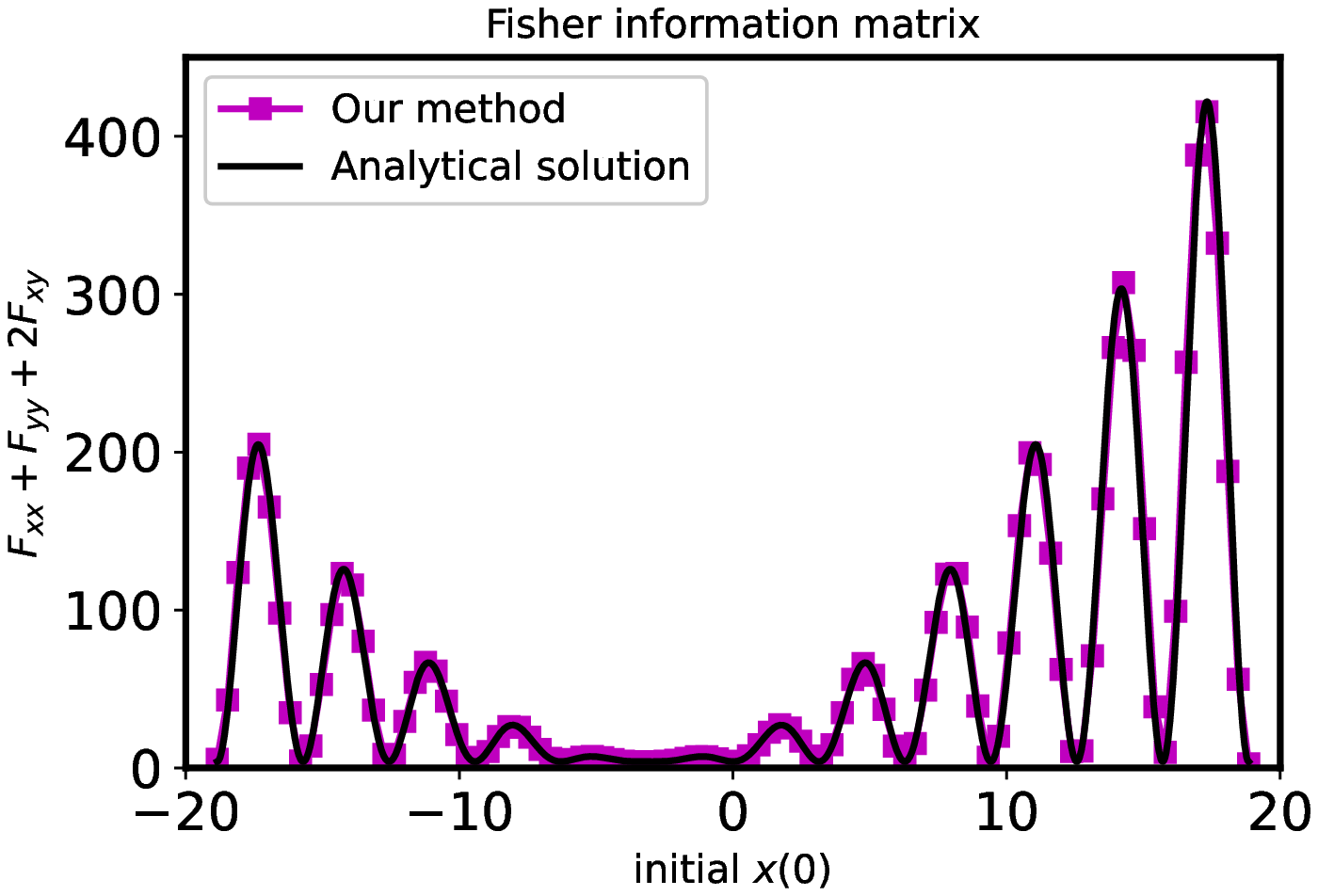}}\\
\subfigure[]{\includegraphics[scale=0.55]{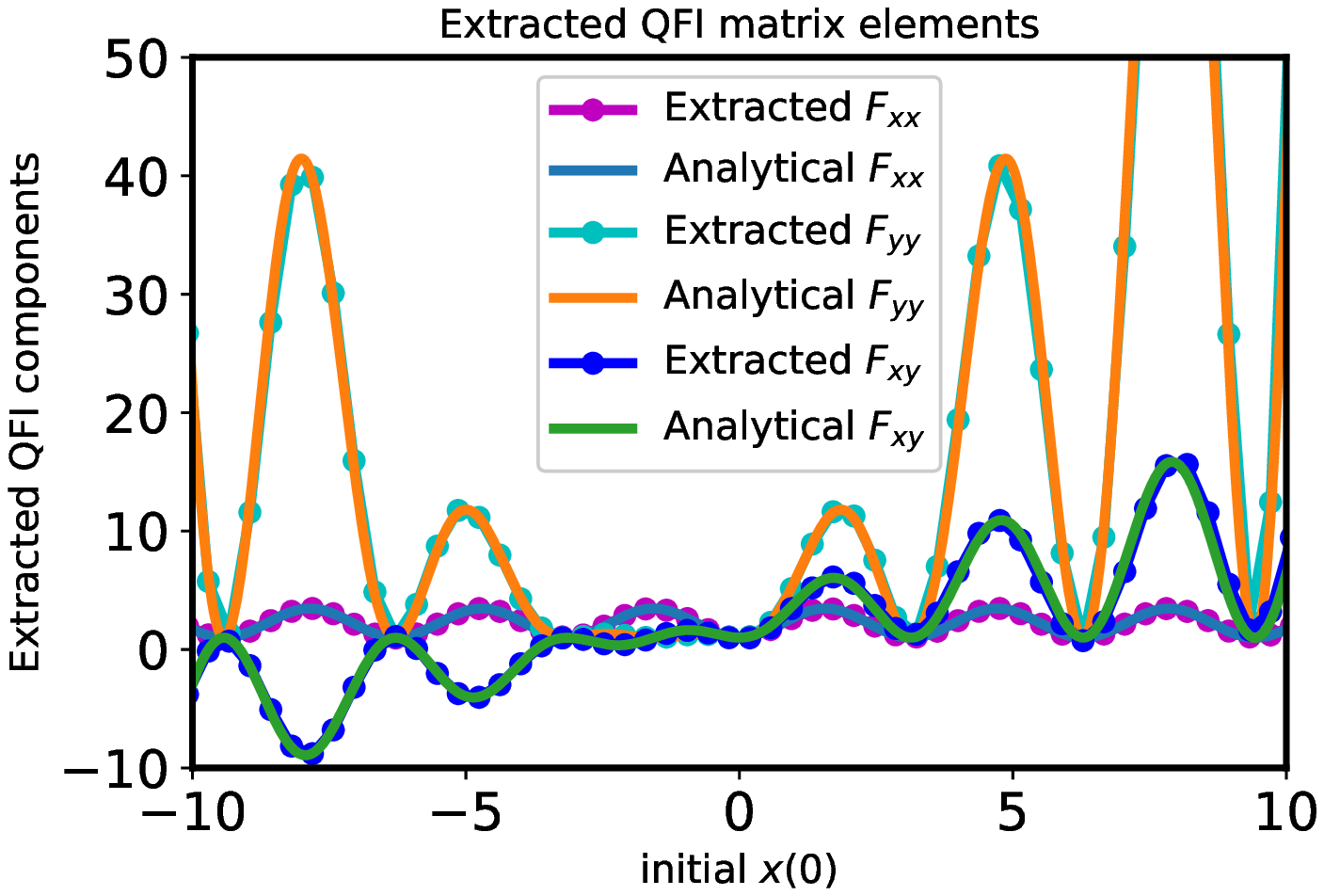}}
\caption{(a) The sum of QFIM with respect to $x(0)$ when the final Hamiltonian's ground state is estimated. (b) The diagonal QFI is evaluated using Eq.(\ref{main}), while the off-diagonal QFI is obtained using half of the difference between the sum of all QFIM elements and all the diagonal QFIM elements.}
\label{off-diagonal QFI}
\end{figure}

In order to validate Eq.~(\ref{secondmain}), we construct a non-trivial two-level example since the Hamiltonian (\ref{twoHam}) with $\theta$ and $\phi$ as parameters has vanishing off-diagonal quantum Fisher information. We still use the form Eq.~(\ref{twoHam}), but the vector becomes $\vec{n}(t)=(\sin(x+y)\cos(xy),\sin(x+y)\sin(xy),\cos(x+y))$, where $x$ and $y$ are the parameters to be estimated. The parameters are driven as followings, $x(t)=x(0)+\frac{v^2 t^2}{2\pi}$ and $y(t)=\frac{v^2 t^2}{2\pi}$. The final time $t_f=\frac{\pi}{v}$ parameters' changing velocities are both $v$. Using our method, we plot the function of the sum of Fisher information matrix elements with respect to the initial $x(0)$. The QFIM is evaluated at the final time Hamiltonian's ground state with respect to $x(t_f)=x(0)+\frac{\pi}{2}$ and $y(t_f)=\frac{\pi}{2}$. From the Fig.~\ref{off-diagonal QFI}, we find our method matches the exact solution perfectly. 

\section{Experimental consideration}
We now carry out some experimental discussions. Here we choose the nitrogen-vacancy (NV) center \citep{DOHERTY20131,PhysRevB.85.205203,Hern_ndez_G_mez_2021,doi:10.1146/annurev-conmatphys-030212-184238,PhysRevLett.124.210502} in diamond as the applicable settings and employ the work of Yu \textit{et al}.\ to exhibit our measuring procedure. The NV center has three sublevels $m_s=0,\pm1$, and using external magnetic field $B_z\simeq510$ Gauss can lift the degeneracy between $|+1\rangle$ and $|-1\rangle$. Then the two lower sublevels $|0\rangle$ and $|-1\rangle$ constitute a two-level system with a gap $\omega_0=D-\gamma_e B_z$, where the $D=2\pi\times2.87$ GHz is the zero-field splitting and $\gamma_e$ is the electronic gyromagnetic ratio. We need to prepare the system in the ground state of the interested Hamiltonian. The standard procedure includes the 532 nm green laser which first initializes the NV center in the $m_s=0$ state. Using an arbitrary waveform generator can drive the transition between levels $|0\rangle$ and $|-1\rangle$, and the Hamiltonian of the laboratory frame is $H=\omega_0\sigma_z /2+V(t)\sigma_x$ \citep{yu2022quantum,10.1093/nsr/nwz193,PhysRevLett.120.120501}. In the rotating frame, the effective Hamiltonian can take the form of $H_{eff} =A/2(\sin\theta\cos\phi\sigma_x+\sin\theta\sin\phi\sigma_y+\cos\theta\sigma_z)$ when $V(t)=A\sin\theta\cos[\omega_0t-f(t)+\phi]$. Applying a microwave field can take the initial state from $m_s=0$ state to the ground state of $H_{eff}$, followed by the adiabatic ramp $\phi=v^2 t^2 /(2\pi)$ with different fixed $\theta$ in different runs of experiment. This kind of ramp has been implemented in Ref.~\citep{PhysRevLett.120.120501}. In the final time $t_f$, the SAH or energy fluctuation operator of the effective Hamiltonian can be measured through fluorescence detection during optical excitation \citep{PhysRevLett.120.120501,Hern_ndez_G_mez_2021,PhysRevResearch.2.023327}.
\begin{figure}[h]
	\centering
	\subfigure[]{\includegraphics[scale=0.6]{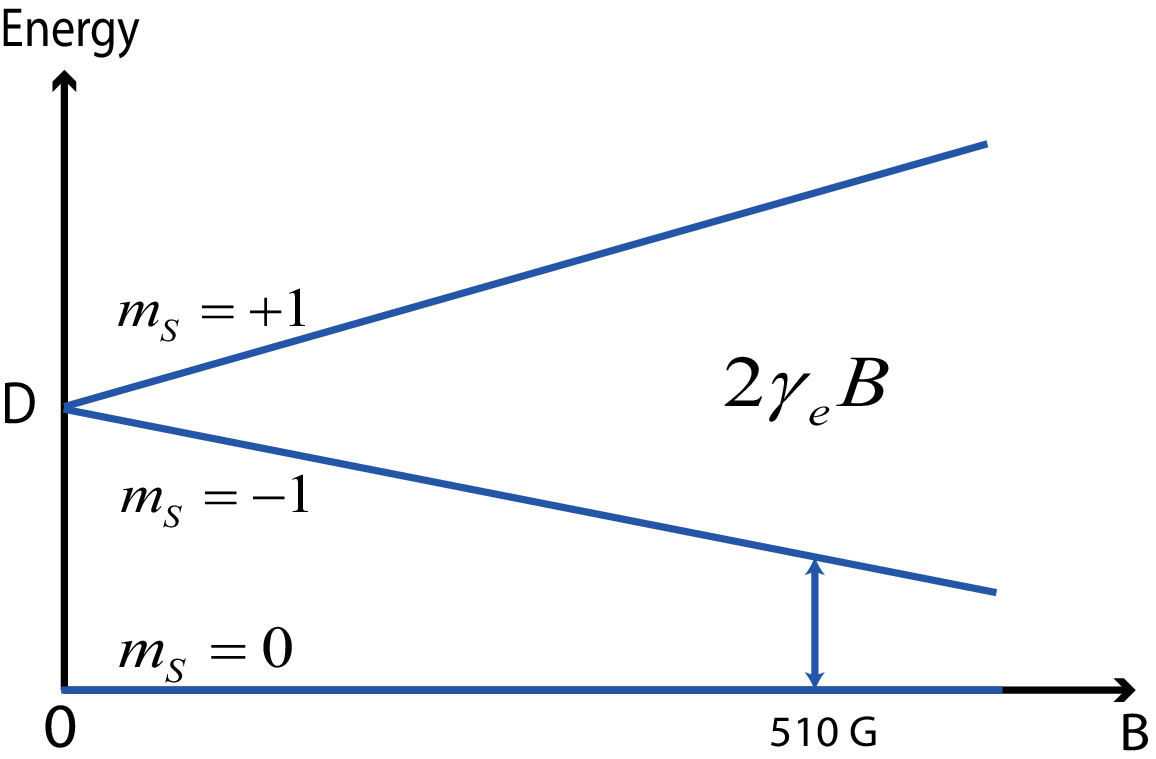}}\\
	\subfigure[]{\includegraphics[scale=0.6]{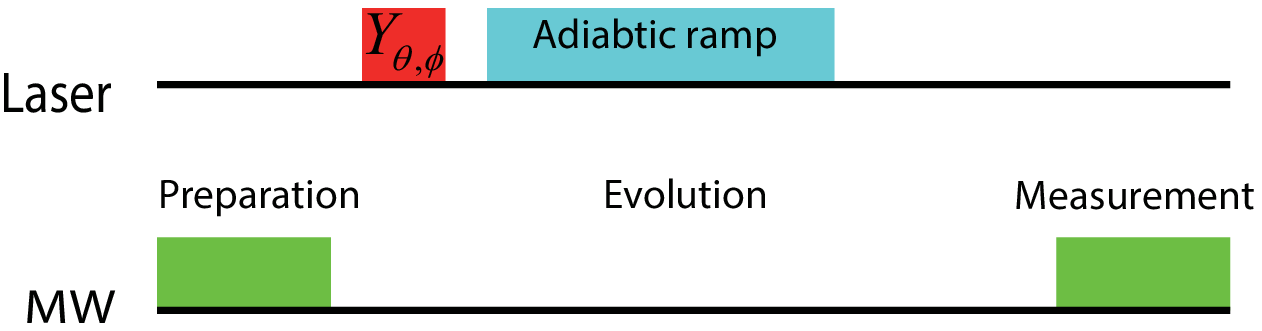}}
	\caption{(a) The energy splitting first due to the zero-field splitting $D$ and then the external field $B$. (b) The demonstration of the protocol of measuring QFI in NV center.}
	\label{NV}
\end{figure}

\begin{figure}[h]
	\includegraphics[scale=0.5]{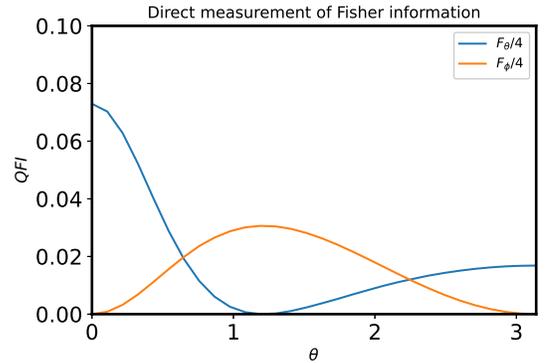}
	\caption{The QFI of $\theta$ and $\phi$ in a rotating frame interacting two-qubit model, which is experimentally measured in \citep{10.1093/nsr/nwz193}. The result using our method with the same parameters is exactly coincident with that of \citep{10.1093/nsr/nwz193}. We list the actual parameters here: $A_x=2.79$ MHz, $A_z=11.832$ MHz, $\Omega_{mw}=2.13$ MHz, $\gamma_n B_{||}=1.07\times749.32$ kHz and $\phi=$ 0.}
	\label{interacting}
\end{figure}

The NV center has been the device for measuring the QFI using the method in Ref.~\citep{PhysRevB.97.201117}. The experimental results have been demonstrated in Ref.~\citep{10.1093/nsr/nwz193}. We utilize our measuring protocol to carry out the numerical simulation with the practical parameters given in Ref.~\citep{10.1093/nsr/nwz193}. The system we take use of is the NV electronic spin coupled by a $^{13}$C nuclear spin, for which the rotating frame interacting two-qubit effective Hamiltonian is given by
\begin{align}
	H_{rot}(\theta,\phi)=\frac{\Omega_{mw}}{2}\times[\cos\theta\sigma_z+\sin\theta(\cos\phi\sigma_z+\sin\phi\sigma_y)]\nonumber\\+\big(\frac{\gamma_nB_{||}}{2}-\frac{A_z}{4}\big)\tau_z-\frac{A_x}{4}\tau_x-\frac{A_z}{4}\sigma_z\otimes\tau_z-\frac{A_x}{4}\sigma_z\otimes\tau_x.\label{interHam}
\end{align}
In Fig.~\ref{interacting}, we give the QFIs $F_\theta$ and $F_\phi$ of the ground state of the Hamiltonian Eq.~(\ref{interHam}) for different values of $\theta$, when $\phi=0$ is set. The simulation result coincides with that in Ref.~\citep{10.1093/nsr/nwz193} well.

\section{Extensions to the statistical models}
The QFI has been considered to be the key to studying the quantum phase transitions. Since our method gives the direct measurement of the QFI, we will
exam the result obtained from the adiabatic perturbation theory and discuss
whether the signal of quantum phase transition is still clear.
\subsection{Transverse field Ising model}
The famous one-dimensional transverse field Ising model is always
the first model to detect new physics under the situation of phase
transition. We study the following model \citep{gu2010fidelity}
\begin{equation}
H(t)=-J\sum_{i=1}^{N}\sigma_{i}^{x}\sigma_{i+1}^{x}-B(t)\sum_{i=1}^{N}\sigma_{i}^{z},
\end{equation}
where $J$ is the coupling strength between neighboring spins and
$B(t)$ is the external magnetic field along the $z$-direction. We fix the
coupling strength $J=10$ and modify the field $B(t)=5+\frac{v^{2}t^{2}}{4h}$
to obtain the QFI, where the field approaches the artificial value
$(5+h)$ at the final time $t_{f}=\frac{2h}{v}$. We can adjust the
evolving speed $v$ and the pre-set value $h$. Through changing the value $h$, we can measure the QFI of the ground state at arbitrary external magnetic field $B(t)$. The initial value
of the magnetic field is set to be $B(0)=5J$ to lift the possible
degeneracy of energy levels. 

\begin{figure}[h]
\includegraphics[scale=0.5]{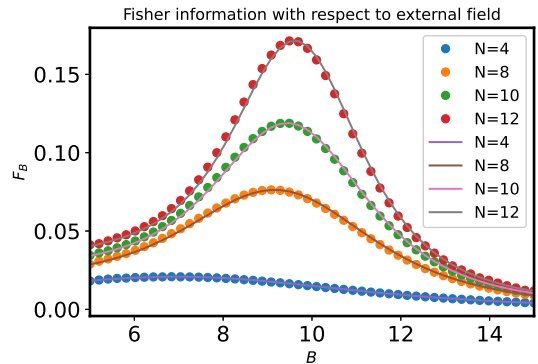}
\caption{The QFI of the ground state of
transverse-field Ising model with respect to the field strength. The dot is the measured quantum Fisher information using our method, while the line is the analytic solution using Eq.~(\ref{FisherAnalIsing}). In our case, the positive parity is used since the energy is lower in the situation of positive parity.}
\label{QFI_TFI}
\end{figure}

The ground state of the Hamiltonian is 
\begin{equation}
|\phi_{0}\rangle=\prod_{k>0}\bigl(\cos\theta_k|0\rangle_k|0\rangle_{-k}+\mathrm{i}\sin\theta_k|1\rangle_k|1\rangle_{-k}\bigr),
\end{equation}
where $|0\rangle_k$ and $|1\rangle_k$ are the number states of the $k$-space fermionic operator before the Bogoliubov transformation. The quantum Fisher information with respect to the external field is evaluated to be \citep{PhysRevLett.99.100603,PhysRevLett.99.095701,PhysRevE.74.031123,PhysRevLett.96.140604}
\begin{equation}
F_B=\sum_{k>0}\frac{J^2\sin^2 k}{\bigl(J^2+B^2-2JB\cos k\bigr)^2}.\label{FisherAnalIsing}
\end{equation}
The peak value of the QFI with respect to the external field should emerge at $B=10$ when $J=10$, corresponding to the phase transition point. Limited by the dimension of the Hilbert space, we give the QFI when the number of spins is $N=4,8,10,12$. We can see from the Fig.~\ref{QFI_TFI} that the peak value becomes gradually distinct when the magnetic field $B=10$, which coincides the theoretical conclusion of this model. It is obvious that the measured quantum Fisher information using our method matches the analytic solution Eq.~(\ref{FisherAnalIsing}) perfectly. 

\subsection{Heisenberg spin chain}
\begin{figure}[h]
	\centering
	\subfigure[]{\includegraphics[scale=0.5]{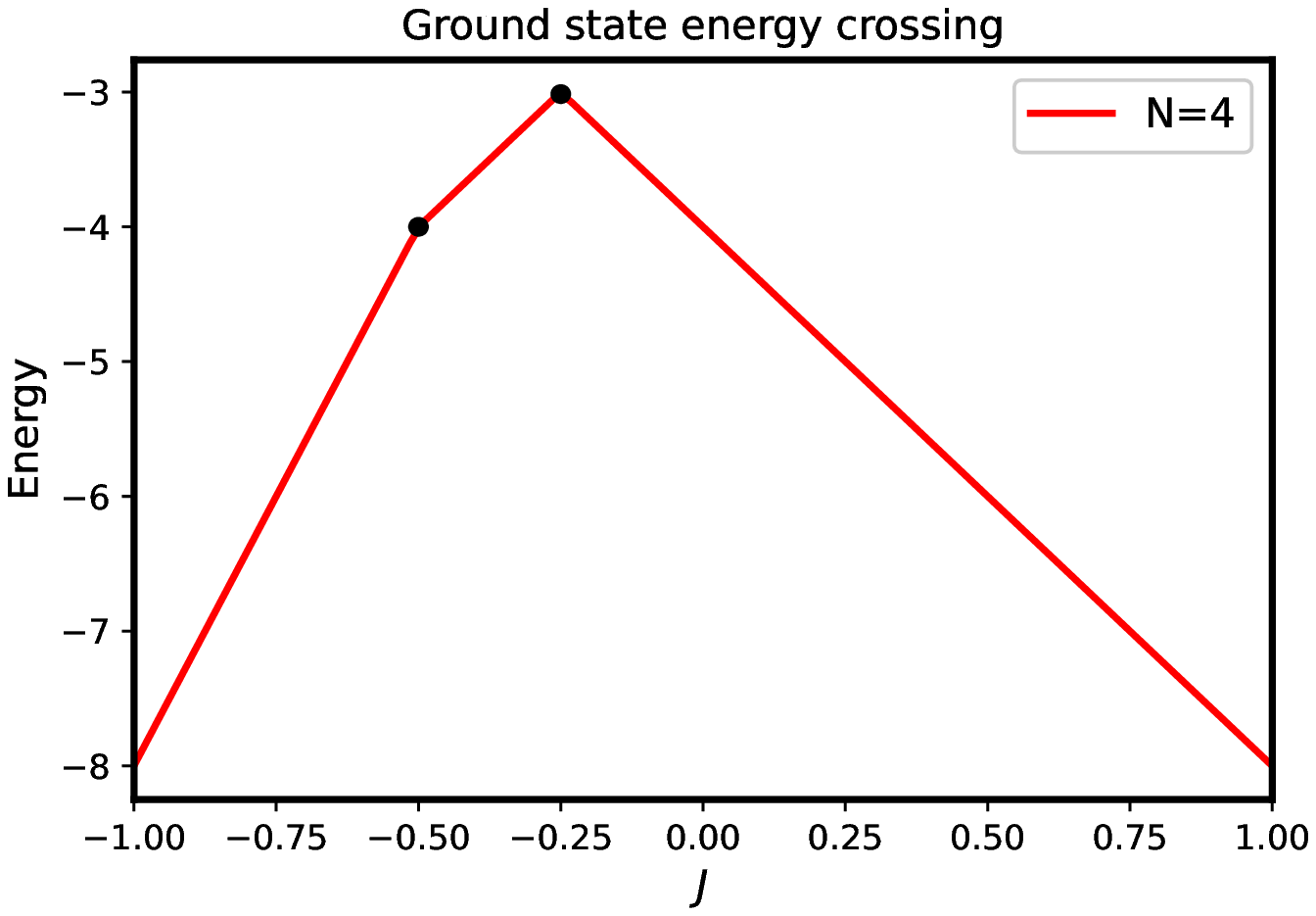}\label{groundstate}}
	\subfigure[]{\includegraphics[scale=0.5]{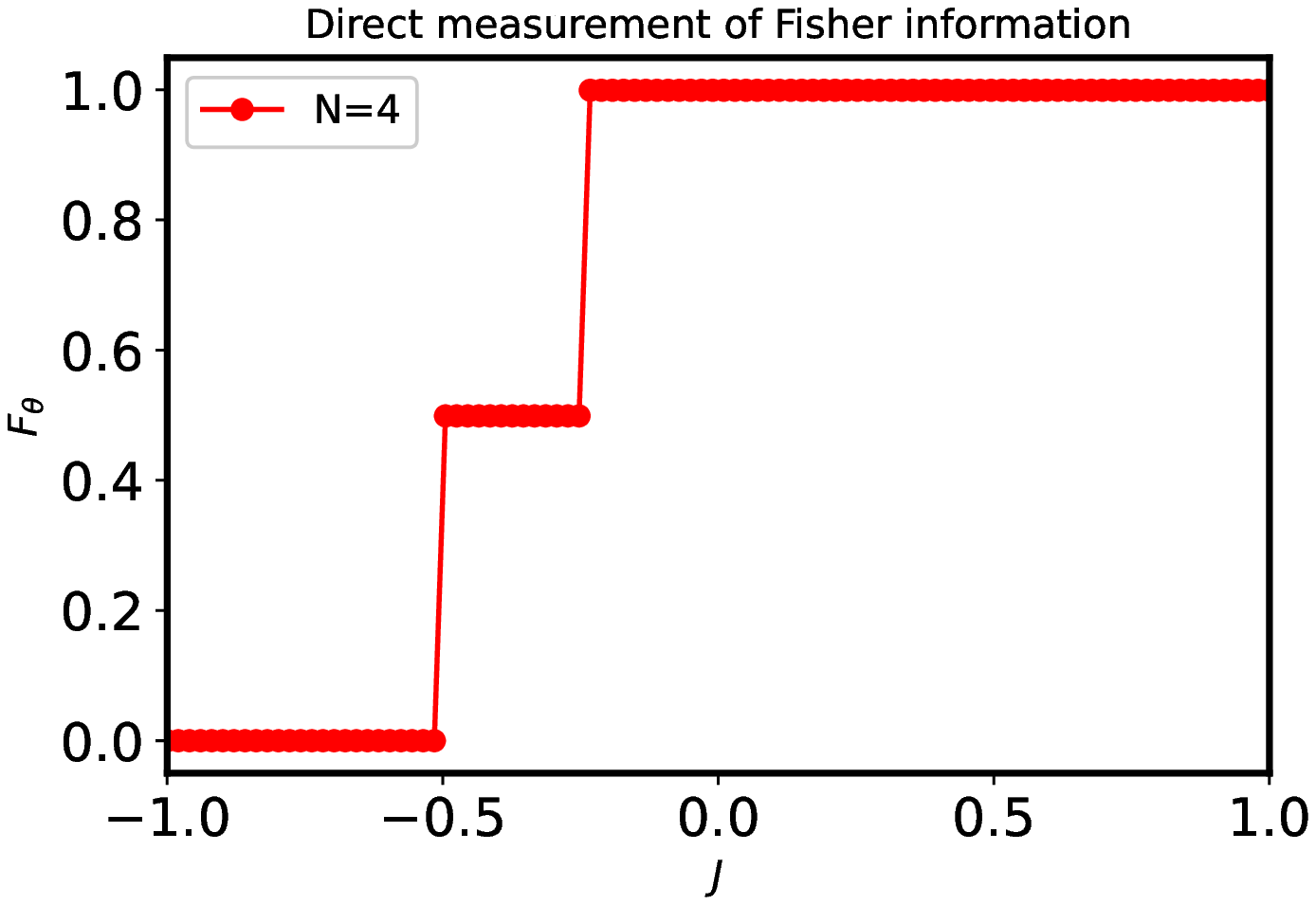}\label{QFI_Heisenberg}}
	\caption{(a) The level crossing of the ground state of Heisenberg spin chain.
		In the considered region of the coupling strength, level crossing
		occurs twice at $J=-0.5$ and $J=-0.25$, respectively. (b) The QFI
		with respect to the field strength of the ground state of Heisenberg
		spin chain. We obtain the QFI using our method. It is apparent that
		the QFI also shows two abrupt changes at the corresponding values
		of $J$, which depicts the level crossing.}
\end{figure}

Here we give another example of application of our measured quantum Fisher information. For the generalized Heisenberg spin chain, when we focus on the coupling strength, the phase transition due to the energy levels crossing emerges under this situation. The Heisenberg spin chain in an external magnetic field \citep{gritsev2012dynamical}
\begin{equation}
	H=-J\sum_{i=1}^{N}\vec{\sigma}_{i}\cdot\vec{\sigma}_{i+1}-\sum_{i=1}^{N}\vec{h}(t)\cdot\vec{\sigma}
\end{equation}
reveals the contribution of level crossing, where $\vec{h}(t)=(\sin\theta(t),0,\cos\theta(t))$
and $\theta=\frac{v^{2}t^{2}}{2\pi}$. The QFI corresponding to the ground state with
respect to $\theta$ at the final time $t=\frac{\pi}{v}$ is obtained
from our method and we plot it as the function of the coupling strength
$J$ in Fig.~\ref{QFI_Heisenberg}.

The final Hamiltonian is actually isotropic spin interaction with a $x$-direction field. The Fig.~\ref{groundstate} denotes the level crossing of the ground state of the final Hamiltonian, and the Fig.~\ref{QFI_Heisenberg} exhibits the step at the corresponding $J$.
Without calculating the exact form of the ground state, we can obtain the fact that the QFI of the ground state with respect to the parameter $\theta$ only changes when the coupling strength $J$ causes the level crossing. The extracted Berry curvature can exhibit such properties as in Ref.~\citep{gritsev2012dynamical}.

\section{Conclusion and Discussion}
In this article, an alternate way is shown to derive the link between SAH or energy fluctuation and the QFI or QFIM. The adiabatic perturbation setup has been employed to measure the Berry curvature via the measurement of the generalized force both numerically and experimentally, hence the extraction of QFI or QFIM can also be applied to the same schemes except the final measured quantity, however, few discussions are carried out. This setup also enables the direct extraction of the quantum geometric tensor. All we need to do is changing the the estimating parameter relatively slowly with zero initial velocity, followed by the measurement of the SAH or energy fluctuation at the final interested moment.

We have adopted two-level systems to testify the measurement of the QFI and QFIM, and both of the extracted values match the analytical results well. A NV center simulation is made using the practical parameters and it fits the experimental results exactly. Since the actual energy gap is large that the parameters' change can be fast enough to fulfill the procedure in the time of magnitude $\mu s$. As a result, the NV center truly can exhibit our scheme. Also, the phase transition and level crossing can also be depicted in this protocol like the Berry curvature. We believe our discussion will make the practical application of the adiabatic perturbation theory in the direct extraction of Berry curvature and QFI, or the full quantum geometric tensor.

\begin{acknowledgments}
This work was supported by the National Key Research and Development
Program of China (Grants No.\ 2017YFA0304202 and No.\ 2017YFA0205700),
the NSFC (Grants No.\ 11875231, No.\ 11935012, No.\ 12175075, and No.\ 61871162), and the Fundamental
Research Funds for the Central Universities through Grant No.\ 2018FZA3005.

\end{acknowledgments}

\end{document}